\documentclass[aps,prl,twocolumn,float]{revtex4}
\usepackage{amsmath,bm,epsfig}

\usepackage[pdftex,
pdfauthor={Oleg Gendelman,Yoav G. Pollack, Itamar Procaccia, Shiladitya Sengupta  and Jacques Zylberg},
pdftitle={Reply to Comment on `What Determines the Static Force Chains in Stressed Granular	Media?'},
pdfsubject={Soft Condensed Matter},
pdfkeywords={Amorphous, force chains, granular, jamming, simulation},
pdfproducer={Latex with hyperref},
pdfcreator={pdflatex}]{hyperref}

\newcommand{\B}[1]{{\bm{#1}}}%% Bold Roman & Greek Lower & Upper Case
\usepackage[dvips]{color}
\begin{document}
\title{Reply to Comment on `What Determines the Static Force Chains in Stressed Granular
Media?'}
\author{Oleg Gendelman$^1$, Yoav G. Pollack$^2$, Itamar Procaccia$^2$, Shiladitya Sengupta$^2$  and Jacques Zylberg$^2$}
\affiliation{$^1$Faculty of Mechanical Engineering, Technion, Haifa 32000, Israel\\ $^2$Dept. of Chemical Physics, The Weizmann Institute of Science,  Rehovot 76100, Israel}
\maketitle

The determination of the normal and tangential forces between frictional disks from visual data
was considered insoluble for three main reasons: (i) the tangential forces that accumulate at contacts
are history-dependent and were believed not to be obtainable from a visual \cite{footnote}, (ii) the number of
mechanical constraints, i.e the vanishing of the net force and the torque on each disk, is much
smaller than the number of inter-disk normal and tangential forces, and the problem is thus under-determined.
(iii) In many realistic granular systems (sand, metallic disks etc.) the compression is so small that the
change in the distances between centers of mass cannot be measured accurately. In the context of an array of disks of
diameters $\sigma_i$,  one can  determine the positions of the center of mass $\B r_i$ relatively easily.
But if the disks are highly incompressible, it is not possible to determined accurately the difference between the
nominal distance $\sigma_i+\sigma_j$ and the actual distance $|\B r_i-\B r_j|$. 
In Ref.~\cite{16GPPSZ} it was shown that given the directions of the vectors connecting the centers of masses
of the disks (but not the actual distances between the center of mass) and the external forces on the disks,
all the normal and tangential forces can be determined {\em exactly} provided the normal forces are linear. There is no
need to know the tangential force law. The solution of all the aforementioned difficulties is achieved by
adding geometric constraints in the form
of the minimal polygons that connect the centers of mass of adjacent disks. 

In a comment on that paper, DeGiuli and McElwaine showed that if the radii of the disks are not
known with proper accuracy,  this
results in errors in the determined forces \cite{16DM}. This is obvious; given highly incompressible disks
in contact, introducing errors in the their radii changes their positions and the vector distances
between the centers of mass. Of course, experimental errors are unavoidable, and care should be
taken to diminish them as much as possible. The theoretical solution of the conceptual difficulties
(i)-(iii) still requires experimental efforts to achieve the highest possible precision. For example
using larger and stiffer disks will automatically reduce the relative error in the radii. The theory in
Ref.~\cite{16GPPSZ} aimed at finding the forces when provided with good measurement of the disk radii; the comment \cite{16DM}
addresses another problem: the statistics of forces in uncertain configuration. The obtained forces
 may depend on inaccuracies, but this fact does not make the solution of Ref.~\cite{16GPPSZ} ``false" in
 any conceivable way as they claim.

Even with experimental errors one can improve the determination of the inter-particle forces
by noticing that the predicted forces do not annul the net force on each particle. An iterative
procedure to achieve such an improvement was proposed in Ref.~\cite{16GPP}. The idea is
to move the particles in the direction of the net force, and recompute the inter-particle
forces. For systems with only normal forces this procedure converges extremely well. An
equivalent procedure for frictional assemblies of disks will be provided elsewhere.

\maketitle


\begin{thebibliography}{99}
\bibitem{footnote}
Photoelastic disks are the only exception where infomration about tangential
forces may be extracted from experiments.
\end{thebibliography}
\end{document}